\documentclass[12pt, draftclsnofoot, onecolumn]{IEEEtran}
\ifCLASSINFOpdf
\else
\fi

\usepackage[subrefformat=parens,labelformat=parens,caption=false,font=footnotesize]{subfig}
\usepackage{array}
\usepackage{amsmath}
\usepackage[nolist,nohyperlinks]{acronym}
\usepackage{graphicx,wrapfig,lipsum}
\usepackage{rotating}
\usepackage{amsmath, amssymb, amsthm}
\usepackage{stmaryrd}
\usepackage{tikz}
\usepackage{verbatim}
\usepackage{url}
\usepackage[utf8]{inputenc}
\usepackage[english]{babel}

\usepackage{multicol}
\usepackage{xr-hyper}

\usepackage{scalerel}
\usepackage{multicol}

\usepackage[noadjust]{cite}
\usepackage{bbm}
\usepackage[normalem]{ulem}
\usepackage{color}
\usepackage{dsfont}
\usepackage{bm}
\usepackage[short]{optidef}
\usepackage{diagbox}
\usepackage{enumitem}
\usepackage{slashbox}
\usepackage[ruled]{algorithm2e}
\setlist{nosep}
\newcommand\blfootnote[1]{%
  \begingroup
  \renewcommand\thefootnote{}\footnote{#1}%
  \addtocounter{footnote}{-1}%
  \endgroup
}

\begin{document}
\title{Optimal Virtual Network Function Deployment for 5G Network Slicing in a Hybrid Cloud Infrastructure}

	\author{Antonio De Domenico$^{1}$, Ya-Feng Liu$^{2}$, and Wei Yu$^{3}$
     \\ \small{ $^1$Huawei Technologies, Paris Research Center, 20 quai du Point du Jour, Boulogne Billancourt,
France; $^2$LSEC, ICMSEC, AMSS,
Chinese Academy of Sciences, Beijing 100190, China; $^3$Department of Electrical and Computer Engineering, University of Toronto, Toronto ON, Canada M5S 3G4}}

\maketitle

        \blfootnote{Part of this work has been presented in IEEE International Conference on Communications 2019 \cite{DeDomenico2018}.        
      This work has been partially performed in the framework of the H2020
project 5G-MoNArch co-funded by the EU. Wei Yu is supported by the Natural Sciences and Engineering Research Council (NSERC) via the Canada Research Chairs Program. Antonio De Domenico was at CEA-LETI, Grenoble,
France, when this manuscript was submitted.}

\begin{abstract}
Network virtualization is a key enabler for 5G systems to support the expected use cases of vertical markets. In this context, we study the joint optimal deployment of Virtual Network Functions (VNFs) and allocation of computational resources in a hybrid cloud infrastructure by taking the requirements of the 5G services and the characteristics of the cloud architecture into consideration. The resulting mixed-integer problem is reformulated as an integer linear problem, which can be solved by using a standard solver. Our results underline the advantages of a hybrid infrastructure over a standard centralized radio access network consisting only of a central cloud, and show that the proposed mechanism to deploy VNF chains leads to high resource utilization efficiency and large gains in terms of the number of supported VNF chains. To deal with the computational complexity of optimizing a large number of clouds and VNF chains, we propose a simple low-complexity heuristic that attempts to find a feasible VNF deployment solution with a limited number of functional splits. Numerical results indicate that the performance of the proposed heuristic is close to the optimal one when the edge clouds are well dimensioned with respect to the computational requirements of the 5G services.
\end{abstract}
%
%
\section{Introduction}
\label{sec:intro}
The development of the fifth generation (5G) system is driven by the goal of addressing new mobile services characterized by heterogeneous requirements, e.g., peak data rate, latency, and network energy efficiency \cite{IMT_5G}. To fulfill this goal, the 5G research community is defining a new flexible architecture, where the network infrastructure is split into logical instances, i.e., network slices, each tailored to a dedicated service and running in a cloud infrastructure.
{A network slice could span across multiple domains,
i.e., radio access network, transport network, and
core network, and could be deployed
across multiple operators. It comprises
dedicated and/or shared resources and can be completely
isolated from the other network slices to fulfill service level agreement \cite{shariat2019flexible}.}

Network slices are composed of a chain of Virtual Network Functions (VNFs), which represent the software implementation of the traditional network functions (NFs), such as coding/encoding or packet scheduling, and can be efficiently reconfigured and optimized through the European Telecommunications Standards Institute (ETSI)
Management and Orchestration (MANO) and Network Function Virtualization (NFV) frameworks \cite{ETSI}.

In the current vision for 5G, depending on the momentary service requirements and network load, the available cloud resources can be dynamically distributed across slices. Moreover, the VNF chain in each slice can be split \cite{Rost2014}, in order to deploy the corresponding VNFs on the proper cloud units, to increase the resource utilization efficiency, or to reduce the end-to-end latency.
However, when implementing such a paradigm it is important to consider that the NFs of traditional mobile communication systems are characterized by tight inter-dependencies \cite{maeder2014}, which are the results of the design assumption that all NFs reside in the same fixed location, e.g., at a base station (BS).
These inter-dependencies result in very stringent latency constraints on the communication link between the Radio Remote Heads (RRHs) and the cloud infrastructure. 

To deal with these constraints and to enable network slicing for 5G {Ultra-Reliable and Low latency Communications (URLLC)} services characterized by low
latency requirements, e.g., augmented reality or factory automation, edge clouds can be deployed
across the network.
{Besides reducing the end-to-end latency, edge clouds limit  backhaul congestion,  enhance privacy, and extend the mobile terminal battery lifetime,  which are key benefits for massive Machine Type
Communications (mMTC) services \cite{ETSIMEC}. Finally, this architecture enables novel mobile services and applications, such as computational offloading or edge artificial intelligence.}
Nevertheless, due to the high cost for site acquisition in urban areas, each
edge  cloud  typically  has  lower  computational  capacity {and storage}  than  a  central  cloud  \cite{EdgeCloud},  which {may lead to poor performance when dealing with big data processing, thus limiting}
the number and types of services that it can support. {In contrast, central clouds can provide high-performance computing services
at the cost of relatively large transmission
latency and stringent backhaul capacity requirements.}
Therefore, in 5G systems, edge cloud and
central cloud facilities {are perfectly paired to accommodate emerging services and} will coexist to form a complex hybrid cloud architecture \cite{Hu2020}.

This calls for cloud orchestration mechanisms that take service requirements  and network constraints jointly into account, to enable efficient and cost-effective 5G service deployment.

\subsection{Related Work}
In the existing literature, the work \cite{Liu2015} investigated the tradeoff between computational and fronthauling costs when deciding the functional split between the RRH and the central cloud. However, it considered cloud units with unlimited capacity. The work \cite{Koutsopoulos} studied the functional split selection and baseband server scheduling problem in a Centralized Radio Access Network (C-RAN). However, the author considered only the overall service latency, without taking into account the requirements of the specific VNFs.
The work \cite{Alabbasi2018} focused on a hybrid C-RAN architecture and investigated the functional split that limits the system power consumption and the bandwidth usage in the link between the edge and the central clouds. Nevertheless, it did not consider the fact that each VNF has processing and latency requirements. 
{The work investigated the problem of the optimal selection of the functional split in small cells with the goal of minimizing inter-cell interference and fronthaul bandwidth utilization \cite{Harutyunyan2018}. The work \cite{Holm2015} studied how to optimize the deployment of cloud nodes for serving different BSs considering jointly statistical multiplexing gain and backhaul cost. Similarly, the work \cite{Carapellese2014} proposed a centralized unit deployment algorithm for optical transport networks,
aiming at minimizing the infrastructure power consumption. However, these works did not focus on network slicing and they did not consider cloud computational capacity.}
The work \cite{Musumeci16} formulated an integer linear programming (ILP) to investigate the optimal deployment of the VNF chain. However, it mainly focused on the fronthaul constraints, without considering the requirements of the network slice and functional splits. The works \cite{Zhang2017} \cite{Zhang2018} considered the VNF deployment problem under computational resource constraints but they did not take the VNF latency requirements into consideration.
The works \cite{Arouk17} and \cite{Arouk2018} investigated the allocation of VNFs in a hybrid cloud infrastructure. They took into account the latency requirements of each VNF; however, they did not consider that functional splits affect the computational resource requirements. { The work \cite{Chen2020} jointly studied the VNF deployment and the associated routing problem to optimize the system energy efficiency.}
All the above works assumed slices with the same constraints; however, 5G systems need to comply with services with diverse requirements, which determine the computational and latency constraints of each VNF. In work \cite{DeDomenico2018}, we investigated the deployment of VNF chains in a simple cloud infrastructure composed by one central cloud and one edge cloud.
\subsection{Contributions}
In this paper, we propose a framework for the optimal deployment of VNF chains of slices with heterogeneous requirements in a hybrid cloud infrastructure with multiple edge clouds and one central cloud. 

Our contributions are summarized as follows:
\begin{itemize}
    \item {In contrast to all previous works, in our analysis, we take into account the type of mobile services associated with each network slice as well as the different requirements of the VNF chains that compose the slices. Accordingly, we model the relations between the VNF computational requirements, their latency constraints, and the cloud characteristics. We highlight that the flexibility brought by a functional split comes at the cost of increased latency, which needs to be compensated by additional computational resources to comply with the VNF requirements. 
    Based on this analysis, we formulate an ILP that can be solved by using a classical solver (e.g., Gurobi \cite{gurobi}).}
    \item {We first study a simple hybrid infrastructure composed of one edge cloud and one central cloud. Our results show that this solution leads to large gains with respect to a standard C-RAN architecture with only a central cloud both in terms of resource usage efficiency and in the number of supported slice chains. In addition, we highlight that the proposed optimal scheme for the VNF deployment outperforms simple static solutions. In fact, our approach is capable of distributing the cloud load (e.g., moving part of a chain from one cloud to another) when the system is saturated, thus making computational resources available where needed by slices with tight requirements.}
    \item {We then extend our framework to the general case of a hybrid cloud infrastructure composed of one central cloud and multiple edge clouds. We show that the new problem is still an ILP. However, the complexity of solving this problem can constrain the system performance, due to a possibly large number of involved variables, and thus, we propose a simple low-complexity heuristic, which carefully exploits the special structures of the problem. Our results show that, when the edge cloud network is well dimensioned with respect to the services with large computational requirements, the proposed heuristic and the optimal solution have similar performance. In contrast, when the edge clouds have limited computational capacity, the optimal solution, which can introduce multiple splits in the VNF chain, leads to large gains in terms of resource utilization efficiency and the number of successfully deployed VNF chains.}
\end{itemize}

The rest of the paper is organized as follows. In Section \ref{sec:format}, we introduce our network slice deployment framework and the associated hybrid cloud infrastructure. In Section \ref{twoclouds}, we describe the VNF deployment problem in the case of a simple two-cloud hybrid architecture; we analyze this problem and reformulate it as an ILP.
Then, in Section \ref{multi-edge}, we extend our analysis to the case of a dense cloud architecture with multiple edge clouds, and we formulate a new ILP, which can be used to optimally solve the deployment problem for this architecture. 
 In Section \ref{heuristic}, we propose a low-complexity heuristic that aims to solve the VNF deployment problem in a sub-optimal way but with a much smaller complexity compared to that of globally solving the problem. Simulation results are provided in Section \ref{sec:simulation}. Finally, the paper concludes in Section \ref{sec:conclusion}. 

\section{Network Slice Deployment in a Hybrid C-RAN}
\label{sec:format}
\begin{figure}[t]
\centering
\captionsetup{justification=centering}
\includegraphics[width=14cm]{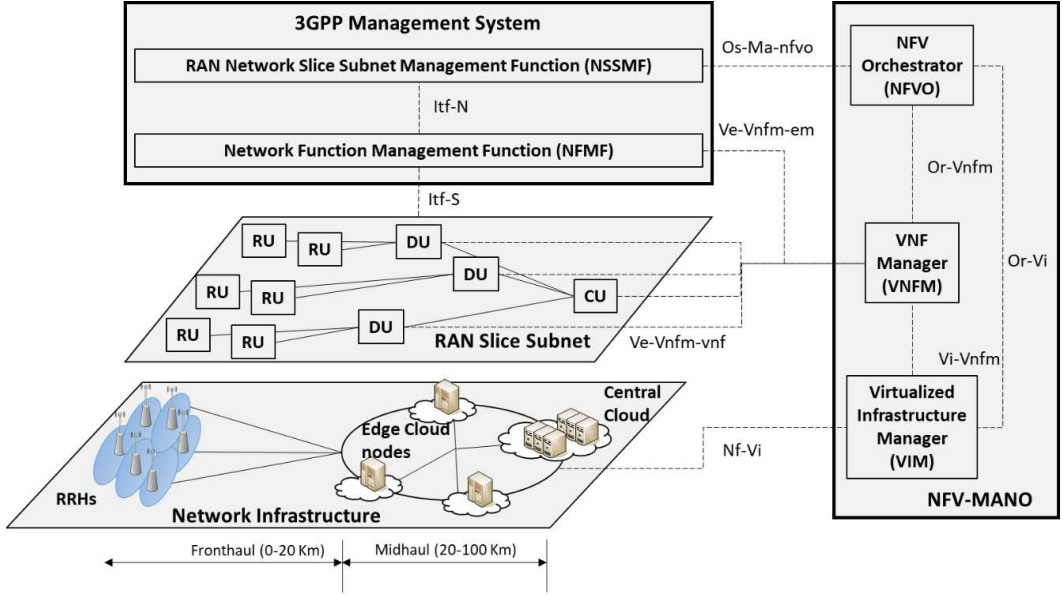}
\caption{Mapping of the hybrid C-RAN deployment to the 3GPP/NFV-MANO management framework \cite{3GPPTS28.533}\cite{Adamuz-Hinojosa2019}.}
\label{fig:SystemModel}
\end{figure}

We consider a C-RAN system supported by a hybrid cloud infrastructure that enables function virtualization and network slicing (see Fig. \ref{fig:SystemModel}). In the Radio Access Network (RAN), a set of RRHs $\mathcal{R}=\left\{1, 2,\ldots,R\right\}$ provides service for applications with different throughput requirements and latency constraints, such as enhanced mobile broadband (eMBB), mMTC, and URLLC. The hybrid cloud is composed of an edge cloud layer, consisting of multiple edge clouds, and one central cloud. We use $\mathcal{K}=\left\{0,1, 2,\ldots,K\right\}$ to denote the set of cloud nodes in the hybrid C-RAN, where the index `$0$' represents the central cloud and $\left\{1, 2,\ldots,K\right\}$ the set of edge clouds. Also, we denote the computational capacity [GFLOPS/s] of the $k$-th cloud as $C^k$, the distance\footnote{Although $d^r_k$ could be a more precise notation, for sake of readability, we drop the index $r$. In fact, in this work we do not focus on the association between the VNF chains and the RRHs, and we assume that each chain is connected to a single RRH $r$ $\in$ $\mathcal{R}$, which yields unique and well-defined $d_k$.} of the $k$-th cloud from the $r$-th RRH as $d_k$, and the distance between two clouds $k,j$ $\in$ $\mathcal{K}$  as $d_{k,j}$. High-capacity, low-latency fiber links characterized by a speed $v$ ($\sim$200 m/$\mu$s) ensure {full direct} connectivity between the RRHs and the clouds.

\begin{figure*}[t]
\centering
\includegraphics*[width=14cm]{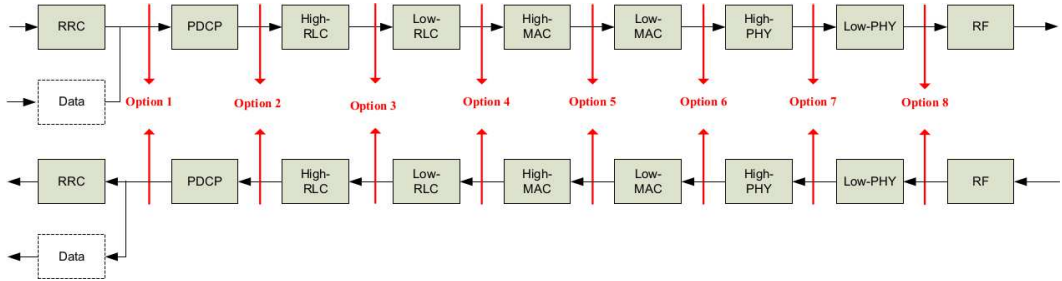}
\caption{3GPP options for the function split between DUs and CUs \cite{3GPPTR38.801}.}
\label{fig:Functional split}
\end{figure*}

{From a functional perspective, 5G BSs are decomposed into radio units (RUs),
distributed units (DUs), and centralized units (CUs) \cite{3GPPTR38.801}, which can be implemented either as VNFs or the standard dedicated hardware (see the RAN slice subnet in Fig. \ref{fig:SystemModel}). The
RUs are the logical entities that correspond to the RRHs, where antennas and Radio Frequency (RF)
hardware are installed. The DUs contain the
low-layer functionalities of the protocol stack whereas the CU includes the high-layer
functionalities. 3GPP has studied eight functional split options \cite{3GPPTR38.801} between CU and DU (see Fig. \ref{fig:Functional split}) but currently the high-layer split is focusing on option 2 while few
variants are under discussion for the lower-layer split \cite{NGMN2019, Adamuz-HinojosaCSCN, Adamuz-Hinojosa2019}.}

{In our framework, mobile services are mapped into a set of network slices\footnote{In the following, we will use the terms slices and services indistinguishably.}, which are associated with one or multiple BSs providing mobile connectivity. As previously mentioned, a 5G BS can be seen as a chain of NFs. Overall, we consider that the network slices require a set of RAN NF} chains $\mathcal{S}=\left\{1,2,\ldots,S\right\}$, each one composed of $N_s$ VNFs. Independently on the type of services, we assume that nine blocks define the NF {chains} as depicted in Fig. \ref{fig:Functional split}: RF, lower physical layer (PHY), higher PHY, lower medium access control (MAC), higher MAC, lower radio link control (RLC), higher RLC, packet data convergence protocol (PDCP), and radio resource control (RRC). The exact content of these blocks depends on the functional split implementation; in our system, {we consider that the eight functional blocks above the RF are all virtualized in the hybrid cloud infrastructure described in the lower part of Fig. \ref{fig:SystemModel}, thus forming a VNF chain. The RF functions are physically implemented in the RRH set $\mathcal{R}$; in addition, depending on the radio (coverage or throughput) requirements, several slices can be supported through the same RRH. Several works have investigated how to properly associate RRHs and network slicing \cite{Afolabi2018}. In contrast,} in this work, we focus on the optimal deployment of VNF chains in a hybrid cloud infrastructure jointly with the associated resource allocation problem in order to satisfy VNF computational and latency constraints.

{This optimization process is carried out in the 3GPP/NFV-MANO network slice management framework described in the upper part of Fig. \ref{fig:SystemModel}. 
To deal with the management and orchestration of the network slice instances, 3GPP decomposes each slice into subnets, each one comprising the NFs of a specific network domain, e.g., access network or core network. In the 3GPP Management System, the Network Slice Subnet Management Function (NSSMF) is responsible for the management and orchestration
of each subnet. Moreover, the Network Function Management Function (NFMF) provides the management services for one or more NFs. However, the cloud architecture management goes beyond the scope of the 3GPP; therefore, the 3GPP management system interacts with the ETSI NFV-MANO to realize the resource management for virtualized core, RAN, and end-to-end network slicing.}

{In sharp contrast with previous works, we assume that each} VNF, depending on the corresponding NF and the associated service, has different latency constraint and computational requirement. Specifically, we use $\lambda_{s,n}$ to indicate the computational requirement [GFLOPS] for $s\in \mathcal{S}, ~n\in \mathcal{N}_s:=\left\{1,2,\ldots,N_s\right\}$. {The computational requirement of a VNF can be determined using a recent empirical model, which can be described by a quadratic polynomial \cite{Khatibi2018}}:
\begin{equation}
    \lambda_{s,n}=\frac{C_{\mbox{\scriptsize exp}}\cdot RB_{s}}{f_{\mbox{\scriptsize CPU}}} \sum_{k=0}^2\left(\alpha_{n,\mbox{\scriptsize DL},k}\cdot i_{s,\mbox{\scriptsize DL}}^k+\alpha_{n,\mbox{\scriptsize UL},k}\cdot i_{s,\mbox{\scriptsize UL}}^k\right).
    \label{comp_req}
\end{equation}
{The model described in \eqref{comp_req} is based on a set of profiling experiments that characterize the complexity of the RAN functions.} {$C_{\mbox{\scriptsize exp}}$ and $f_{\mbox{\scriptsize CPU}}$ are the computational capacity [GFLOPS/s] and the frequency of the CPU [GHz] of the machine used for those experiments \cite{Khatibi2018}}, $RB_{s}$ is the number of resource blocks (RBs) allocated to {NF chain} $s$, $i_{s,\mbox{\scriptsize DL}}$ and $i_{s,\mbox{\scriptsize UL}}$ are the indices of the modulation and coding schemes (MCSs), as defined in the MCS table in 3GPP TS 38.214 \cite{3GPPTS38.214}, for chain $s$ in the downlink (DL) and the uplink (UL), respectively: the higher the index, the higher the MCS spectral efficiency. Moreover, $\left\{\alpha_{n,\mbox{\scriptsize DL},k},\alpha_{n,\mbox{\scriptsize UL},k}\right\}_{n,k}$ are {the} fitting coefficients {of the quadratic polynomial, which describes the computational requirement of a specific VNF, i.e., different sets of coefficients are used for distinct VNFs in the same slice chain}.

It is well understood that the VNFs of the PHY layer, especially the encoding/decoding, are the most computational demanding functions in the cloud \cite{Rost2014}. {For a given VNF chain, the larger computational demand of lower layer VNFs with respect to functions related to upper layers is modelled by fitting coefficients in \eqref{comp_req} with larger values \cite{ Khatibi2018}.}
In addition, from \eqref{comp_req}, the computational requirement of a VNF increases with the spectral efficiency and the number of RBs, which together define the average  throughput requirement of the slice chain; therefore, applications characterized by very high data rate, such as virtual/augmented reality, are more computational demanding than classical services such as eMBB or mMTC.

The latency constraints of each VNF are defined by the interactions in the NF chain: in particular, VNF $n$ receives inputs from VNF $n-1$, passes its outputs to VNF $n+1$, and potentially provides a feedback to VNF $n-1$, e.g., an (positive/negative) acknowledgment. This process is characterized by timing requirements that guarantee reliable operations in the mobile service \cite{maeder2014}. In this work, we denote as $f_{s,n}$ and $b_{s,n}$ the latency constraints of VNF $n \in \mathcal{N}_s$ with respect to the forward VNF $n+1$ and the backward VNF $n-1$. Among these requirements, the most stringent ones are related to the slot length at the PHY layer (from 31.25 $\mu$s to 1 ms) \cite{3GPPTS38.202} and the hybrid automatic repeat request (HARQ) feedback (8 ms) at the MAC layer \cite{maeder2014}. In addition, depending on the service, these requirements may be looser or more stringent. For instance, the time slot length depends on the service latency requirements, e.g., the shorter slot length format is likely used for URLLC slices. Moreover, some functions may not be deployed in all of the slices, e.g., the HARQ may not be implemented in mMTC slices, which does not require high reliability, or in URLLC services like factory automation, where the overall RAN latency has to be in the order of 1 ms \cite{andrews2014will}.

A virtualized communication system requires the sum of the processing and communication delays to be below these latency constraints. Then, for each $s\in \mathcal{S}$, $n \in \mathcal{N}_s$, we have
\begin{equation}
\label{eq:lat_const}
 \begin{array}{l}
l_{s,n}^{\mbox{\scriptsize P}}+f_{s,n}^{\mbox{\scriptsize C}} \leq f_{s,n} \mbox{  and  }
l_{s,n}^{\mbox{\scriptsize P}}+b_{s,n}^{\mbox{\scriptsize C}}\leq b_{s,n},
\end{array}
\end{equation}
where $l_{s,n}^{\mbox{\scriptsize P}}$ is the processing latency related to VNF $n$ and $f_{s,n}^{\mbox{\scriptsize C}}$ and $b_{s,n}^{\mbox{\scriptsize C}}$ are the latencies related to the communication with the neighbouring VNFs $n+1$ and $n-1$, respectively. 
The processing latency $l_{s,n}^{\mbox{\scriptsize P}}$ depends on the computational requirement and the computational rate [GFLOPS/s] allocated to the VNF.
Thus, to decrease the processing delay of VNFs with tight constraints, a resource orchestrator may choose to deploy them at the central cloud where more computational resources are available. However, this choice increases the communication latencies ($f_{s,n}^{\mbox{\scriptsize C}}$ and $b_{s,n}^{\mbox{\scriptsize C}}$), since the central cloud is typically located in a remote area, far from the access network. This highlights a tradeoff between the usage of central and edge cloud resources and calls for a carefully designed procedure that takes into account the limited computational resources as well as the latencies introduced by the link between the central cloud and the edge cloud, and between them and the access network.

To conclude, in order to maximize the system performance and the number of {VNF chains} that can be deployed, each chain can be further split within the cloud infrastructure, and its components deployed in the most appropriate clouds, in terms of resource availability and VNF requirements. Specifically, multiple splits can be realized within each chain, to increase the system flexibility. On the one hand, this enables improvement in the resource utilization efficiency; on the other hand, it results in increasing the latency experienced by the VNFs of the slice, which is distributed among different clouds. One objective of this work is to understand whether introducing multiple functional splits in a VNF chain brings significant advantages, and under which conditions.

In the next sections, to deal with the optimal usage of the available computational resources, we first investigate the optimal joint computational resource allocation and VNF deployment in a simple two-cloud hybrid architecture, composed of one edge cloud and one central cloud. Then, we extend the problem, considering a hybrid infrastructure with multiple edge clouds and one central node.

\section{Optimal resource allocation and VNF deployment in a two-cloud hybrid infrastructure}
\label{twoclouds}
\subsection{Problem Formulation}
 We formulate the problem of minimizing the total computational resources needed to run $S$ VNF chains by optimizing the VNF deployment and the resource allocation in a two-cloud hybrid infrastructure, as follows:
\begin{mini!}|s|[2]
{\left\{R_{s,n}^c,R_{s,n}^e,x_{s,n}\right\}}{\sum\limits_{s \in \mathcal{S}}\sum\limits_{n \in \mathcal{N}_s} \left(R_{s,n}^c x_{s,n}+R_{s,n}^e(1-x_{s,n})\right)}{}{}\label{P:dim}
\addConstraint{\sum\limits_{s \in \mathcal{S}}\sum\limits_{n \in \mathcal{N}_s} R_{s,n}^cx_{s,n}}{\leq C^c}{}\label{P:dim_con3}
\addConstraint{\sum\limits_{s \in \mathcal{S}}\sum\limits_{n \in \mathcal{N}_s} R_{s,n}^e(1-x_{s,n})}{\leq C^e}{}\label{P:dim_con4}
\addConstraint{x_{s,n}}{\in \{0,1\},\; s \in \mathcal{S}, n \in \mathcal{N}_s}{}\label{P:dim_con5}
\addConstraint{l_{s,n}^{\mbox{\scriptsize P}}+f_{s,n}^{\mbox{\scriptsize C}}}{\leq f_{s,n}, \; s \in \mathcal{S}, n \in \mathcal{N}_s}{
}\label{P:dim_con1}
\addConstraint{l_{s,n}^{\mbox{\scriptsize P}}+b_{s,n}^{\mbox{\scriptsize C}}}{\leq b_{s,n}, \; s \in \mathcal{S},  n \in \mathcal{N}_s}{},\label{P:dim_con2}
 \end{mini!}
 where 
 \begin{itemize}
     \item {
$R_{s,n}^c$ and $R_{s,n}^e$ denote respectively the computational rates allocated by the central cloud and the edge cloud to VNF $n$ of chain $s$;}
     \item {
 $x_{s,n}$ is a binary indicator variable with $x_{s,n}=1$ if VNF $n$ runs at the central cloud and $x_{s,n}=0$ if it runs at the edge cloud\footnote{Note that, to analyze the architecture with multiple edge clouds in Section \ref{multi-edge}, we use a more general notation, i.e., $x^k_{s,n}$=1 when VNF $n$ of {chain} $s$ is deployed at cloud $k$, and zero otherwise. Although the notations used in the two sections are slightly different, they can be easily mapped to each other.}.}
  \end{itemize}
We use $C^e$ and $C^c$ to indicate the computational capacities at the edge cloud and the central cloud, respectively, and \eqref{P:dim_con3} and \eqref{P:dim_con4}
denote the resource allocation constraints at the central cloud and the edge cloud, respectively.
Also, \eqref{P:dim_con1}-\eqref{P:dim_con2} represent the latency constraints related to the interactions between VNFs $n$, $n-1$, and $n+1$, as described in \eqref{eq:lat_const}.
Specifically, the processing latency $l_{s,n}^{\mbox{\scriptsize P}}$ depends on the cloud where VNF $n$ runs and the computational rate it receives, and can be modelled as follows:
\begin{equation}
\label{eq:lat-C}
l_{s,n}^{\mbox{\scriptsize P}}=\frac{\lambda_{s,n}x_{s,n}}{R_{s,n}^c}+\frac{\lambda_{s,n}(1-x_{s,n})}{R_{s,n}^e},\; s \in \mathcal{S}, n \in \mathcal{N}_s,
\end{equation}
where $\lambda_{s,n}$ is defined in \eqref{comp_req}.
The first term and the second term in \eqref{eq:lat-C} represent the latency experienced by VNF $n$ if it is executed at the central cloud or at the edge cloud,  respectively.

{It is worth to highlight that in problem (3) there is no explicit constraint that avoids ping-pong on the VNF deployment, leading communication to flow back and forth between cloud nodes, which affects the transport network load. Note that, in a transport network with limited capacity, increasing the traffic load would inflate the communication latency between the two links, which needs to be compensated by a larger amount of computational resources. Therefore, such a solution would not minimize the objective in \eqref{P:dim}.}

Concerning the communication latency, we assume it to be negligible when the related VNFs are located in the same cloud.
In contrast, if there is a split between $n$ and its backward neighbor $n-1$, the related latency $b_{s,n}^{\mbox{\scriptsize C}}$ is computed as follows:
\begin{equation}
b_{s,n}^{\mbox{\scriptsize C}} = 
    \begin{array}{l}
\begin{cases}
\frac{d_{e,c}}{v}\left|x_{s,n} - x_{s,n-1}\right|, &  \mbox{if } s \in \mathcal{S}, n \in \mathcal{N}_s\backslash \{1\}; \\
  \frac{1}{v}\left(x_{s,1}d_{c}+(1-x_{s,1})d_{e}\right),  & \mbox{if } s \in \mathcal{S}, n=1, 
  \end{cases} \label{eq:com lat}
    \end{array}
\end{equation}
where $d_{e,c}$ denotes the distance between the edge cloud and the central cloud and $d_{e}$ ($d_{c}$) represents the distance of the edge cloud (central cloud) from the RRH where the RF function of NF chain $s$ is deployed and $v$ is the speed of the fiber link used to connect them.
The first case in \eqref{eq:com lat} describes the communication delay for VNF $n \in \mathcal{N}_s\backslash \{1\}$, when 1) VNF $n$ is at the central cloud and VNF $n-1$ is at the edge cloud, 2) VNF $n$ is at the edge cloud and VNF $n-1$ is at the central cloud, and 3) VNFs $n$ and $n-1$ are at the same cloud.
The second case describes the communication latency of VNF $1$ with respect to the NFs that cannot be virtualized (i.e., the RF functions must be implemented in the RRHs). This latency depends on whether VNF $1$ runs at the central or at the edge cloud{, it is always greater than zero, and contributes to the end-to-end latency experienced in the VNF chain.}

Similarly, if there is a split between VNF $n$ and its forward neighbor VNF $n+1$, the communication delay $f_{s,n}^{\mbox{\scriptsize C}}$ is computed as follows:
\begin{equation}
f_{s,n}^{\mbox{\scriptsize C}} = 
\begin{array}{l}
\begin{cases}
\frac{d_{e,c}}{v}\left|x_{s,n} - x_{s,n+1}\right|,  &\mbox{if } s \in \mathcal{S}, n \in \mathcal{N}_s \backslash \{N_s\}; \\
  0, & \mbox{if } s \in \mathcal{S}, n=N_s. 
  \end{cases} \label{eq:com lat +1}
    \end{array} 
\end{equation}
The first case in \eqref{eq:com lat +1} describes the communication delay for VNF $n \in \mathcal{N}_s\backslash \{N_s\}$, when 1) VNF $n$ is at the central cloud and VNF $n+1$ is at the edge cloud, 2) VNF $n$ is at the edge cloud and VNF $n+1$ is at the central cloud, and 3) VNFs $n$ and $n+1$ are at the same cloud.
The second case indicates that there is no forward delay related to VNF $N_s$ as it is the last VNF in the chain. Therefore, for VNF $N_s$, \eqref{P:dim_con1} becomes as follows:
\begin{equation*}
l_{s,N_s}^{\mbox{\scriptsize P}}\leq f_{s,N_s},
\end{equation*}
i.e., the forward latency constraint for VNF $N_s$ guarantees that the overall VNF chain is completed on time and takes only the processing latency of VNF $N_s$ into account.

Problem (3) is a mixed-integer problem with $3\sum\limits_{s \in \mathcal{S}}N_s$ variables. {In general, finding its optimal solution may require an exhaustive enumeration, i.e., its worst-case complexity increases
exponentially with the total number of variables}; however, by taking advantage of its special structures, i.e., the per VNF latency constraints, in the next section, we reformulate it as an ILP {with fewer variables}, which can be optimally solved through numerical solvers (as Gurobi \cite{gurobi}), with a lower complexity. {Note that the linearity of the reformulated problem is also vital as it allows to leverage efficient integer programming solvers to achieve global optimality.}  

\subsection{Analysis and Reformulation}
In this subsection, we derive the {minimum} computational resources needed for satisfying each VNF latency requirement, by fixing the associated (central/edge) cloud nodes. Then, using this, we reformulate the mixed-integer problem $(3)$ as an ILP, thereby allowing the optimization of the deployment of the VNFs on the (central/edge) cloud nodes. 

\subsubsection{Modeling the minimum required computational rates}
\label{Analysis}
The minimum computational rate required by a VNF to satisfy its latency and computational requirements depends on whether or not it is at the same cloud as its neighbouring VNFs.
In general, deploying the entire chain at the same cloud reduces the slice resource footprint, thus increasing the number of services that can be supported by the cloud architecture. 
Specifically, for any $s \in \mathcal{S}$, the computational rate needed by VNF $n\in\mathcal{N}_s\setminus \left\{1,N_s\right\}$, if it is co-located with VNFs $n+1$ and $n-1$ is as follows:
 $$C_{s,n} = \frac{\lambda_{s,n}}{\min\left\{f_{s,n}, b_{s,n}\right\}}.$$

{It is worth to highlight that $\left(R_{s,n}^c x_{s,n}+R_{s,n}^e(1-x_{s,n})\right) \geq C_{s,n}$, i.e., the computational rate allocated by the cloud nodes to VNF $n$ has to be larger than or equal to  the ratio between the computational requirement $\lambda_{s,n}$ of VNF $n$ and the most stringent of its latency constraints $f_{s,n}$ and $b_{s,n}$.} 
However, it may be necessary to make a split in the VNF chain to 1) bring VNFs with tighter latency constraints closer to the RRHs, 2) deploy VNFs with high computational requirements at the central cloud, 3) to make room at a given cloud for additional VNF chains, and 4) balance the computational load in different clouds. 
In this case, {the computational rate allocated by the cloud nodes to VNF $n$ may be strictly larger than $C_{s,n}$. If} VNFs $n$ and $n+1$ are not at the same node, the computational rate required by VNF $n$ may need to be increased to compensate for the \textit{forward} communication delay introduced by the functional split; accordingly, we have
$$C_{s,n}^{+}=\max\left\{\frac{\lambda_{s,n}}{f_{s,n}-\frac{d_{e,c}}{v}},C_{s,n}\right\},$$  with the constraint $d_{e,c}<v\cdot f_{s,n}$, i.e., it is not possible to split VNFs $n$ and $n+1$ if the distance between the central cloud and the edge cloud is larger than $v\cdot f_{s,n}$.
Similarly, if VNFs $n$ and $n-1$ are not at the same node, due to the \textit{backward} communication delay, the needed computational rate is
$$C_{s,n}^{-}=\max\left\{\frac{\lambda_{s,n}}{b_{s,n}-\frac{ d_{e,c}}{v}},C_{s,n}\right\},$$
with the constraint 
$d_{e,c}<v\cdot b_{s,n}$, i.e.,
the distance between the two clouds also limits the possible split between VNFs $n$ and $n-1$ due to the related communication latency. 

Therefore, when optimizing the slice deployment, the forward and backward functional split constraints must be considered, and they can {respectively} be written as follows\footnote{Note that we have reformulated the above strict inequalities as non-strict ones to guarantee that the corresponding sets are closed. When an equality holds, an infinite amount of resources is required to satisfy the VNF latency constraint; however, this will not be selected as a feasible solution due to the computational resource constraints \eqref{P:dim_con3} and \eqref{P:dim_con4}.}:
\begin{equation}
\label{eq:forward}
 \begin{array}{ll}
&d_{e,c}\left|x_{s,n} - x_{s,n+1}\right|\leq v\cdot f_{s,n} \mbox{ and }
d_{e,c}\left|x_{s,n} - x_{s,n-1}\right|\leq v \cdot b_{s,n}, 
 \end{array}
\end{equation}
where $\left|x_{s,n} - x_{s,n+1}\right|$ and $\left|x_{s,n} - x_{s,n-1}\right|$ are equal to one if there is a functional split between VNF $n$ and its respective forward neighbour VNF $n+1$ and backward neighbour VNF $n-1$, and zero otherwise.
Here \eqref{eq:forward} highlights why the optimal slice deployment is a challenging combinatorial problem, in particular when dealing with services with diverse constraints.
Now, let us define for each $s \in \mathcal{S}, n\in\mathcal{N}_s\setminus \left\{1, N_s\right\}$,
$$\Delta C_{s,n}^{-} =C_{s,n}^{-} - C_{s,n}; ~\Delta x_{s,n}^{c-}=\max\left\{x_{s,n}-x_{s,n-1}, 0\right\} ; ~\Delta x_{s,n}^{e-}=\max\left\{-x_{s,n}+x_{s,n-1}, 0\right\},$$ 
$$\Delta C_{s,n}^{+} =C_{s,n}^{+} - C_{s,n}; ~\Delta x_{s,n}^{c+}=\max\left\{x_{s,n}-x_{s,n+1}, 0\right\}; ~\Delta x_{s,n}^{e+}=\max\left\{-x_{s,n}+x_{s,n+1}, 0\right\},$$
where $\Delta C_{s,n}^{-}$ and $\Delta C_{s,n}^{+}$ describe the additional rates, with respect to $C_{s,n}$, required by VNF $n$ when respectively VNF $n-1$ and VNF $n+1$ are at a different cloud; $\Delta x_{s,n}^{c-}$  and $\Delta x_{s,n}^{c+}$ indicate whether there is a split between VNF $n$ located at the central cloud and its neighbouring VNFs $n-1$ and $n+1$, respectively; $\Delta x_{s,n}^{e-}$ and $\Delta x_{s,n}^{e+}$ denote whether there is a split between VNF $n$ located at the edge cloud and its neighbouring VNFs $n-1$ and $n+1,$ respectively.

Using the above notations, for VNF $n\in\mathcal{N}_s\setminus \left\{1,N_s\right\}$, the computational rate to be allocated at the central cloud $\bar{R}^{c}_{s,n}$ or at the edge cloud $\bar{R}^{e}_{s,n}$ can be computed as follows:
\begin{equation}
\label{comp_res}
 \begin{array}{ll}
     \bar{R}^{c}_{s,n}=C_{s,n} x_{s,n}
+ \max\left\{\Delta C^{-}_{s,n} \Delta x^{c-}_{s,n},\Delta C^{+}_{s,n} \Delta x^{c+}_{s,n}\right\};\\
\bar{R}^{e}_{s,n}
= C_{s,n} (1-x_{s,n})
     + \max\left\{\Delta C^{-}_{s,n} \Delta x^{e-}_{s,n}, \Delta C^{+}_{s,n} \Delta x^{e+}_{s,n}\right\}.
\end{array}
\end{equation}
In the above expressions, the first term corresponds to the minimum computational rate required by VNF $n$ when it is co-located with VNFs $n+1$ and $n-1$; moreover, the second term denotes the additional computational rate required in case of the functional split. Note that, when both VNFs $n+1$ and $n-1$ are processed at a different cloud from VNF $n$, the additional resources needed depend on the most stringent constraint between $f_{s,n}$ and $b_{s,n}$ and this is why there is a $\max$ operator in \eqref{comp_res}.

The derivations of the minimum computational rates for the first and last VNFs in each chain are special cases of the previous analysis. 
In particular, considering that for VNF $N_s$ there may exist a split only with VNF $N_s-1$, the computational rate required by $N_s$ when $N_s-1$ is at the same cloud is as follows:
   $$C_{s,N_s} = \frac{\lambda_{s,N_s}}{\min\left\{f_{s,N_s}, b_{s,N_s}\right\}}.$$ If $d_{e,c}<v\cdot b_{s,N_s}$, VNFs $N_s$ and $N_s-1$ may not run at the same cloud; in this case, the computational rate required to process $N_s$ can be computed as
   $$C_{s,N_s}^{-}=\max\left\{\frac{\lambda_{s,N_s}}{b_{s,N_s}-\frac{d_{e,c}}{v}},C_{s,N_s}\right\}.$$
Let us denote as $\Delta C_{s,N_s}^{-} =C_{s,N_s}^{-} - C_{s,N_s}$ the additional rate, required by VNF $N_s$ when VNF $N_s-1$ is at a different cloud; then, the overall computational rate that VNF $N_s$ requires at the central cloud or at the edge cloud is as follows:
\begin{equation}
\begin{array}{ll}
\bar{R}^{\mbox{\scriptsize c}}_{s,N_s}=C_{s,N_s}x_{s,N_s}+ \Delta C^{-}_{s,N_s}  \Delta x^{c-}_{s,N_s};\\
\bar{R}^{\mbox{\scriptsize e}}_{s,N_s}=C_{s,N_s}(1-x_{s,N_s}) +\Delta C^{-}_{s,N_s}  \Delta x^{e-}_{s,N_s},
\end{array}
\end{equation}
where $\Delta x_{s,N_s}^{c-}=\max\left\{x_{s,N_s}-x_{s,N_s-1}, 0\right\}$ and $\Delta x_{s,N_s}^{e-}=\max\left\{-x_{s,N_s}+x_{s,N_s-1}, 0\right\}$ indicate if there is a split between VNF $N_s$, respectively located at the central or at the edge cloud, and VNF $N_s-1$. 

For VNF $1$, in contrast to the other VNFs, the minimum needed computing resources depends on whether it runs at the edge or central cloud, even if it is co-located with VNF $2$.
When VNFs $1$ and $2$ run at the central cloud, the computational rate required for VNF $1$ is as follows: 
\begin{equation*}
C_{s,1}^c=\max\left\{\frac{\lambda_{s,1}}{f_{s,1}},\frac{\lambda_{s,1}}{b_{s,1}-\frac{ d_{c}}{v}}\right\},
\end{equation*}
which highlights that VNF $1$ can run at the central cloud only if $
d_{c}<v\cdot b_{s,1}$.
Moreover, when VNFs $1$ and $2$ are processed at the edge cloud, the computational rate required for VNF $1$ is as follows:
\begin{equation*} 
C_{s,1}^e=\max\left\{\frac{\lambda_{s,1}}{f_{s,1}},\frac{\lambda_{s,1}}{b_{s,1}-\frac{d_{e}}{v}}\right\},
\end{equation*} 
where VNF $1$ can be deployed at the edge cloud only if $d_{e}<v\cdot b_{s,1}$. These constraints on the deployment of VNF $1$ can be explicitly written as follows\footnote{As in \eqref{eq:forward}, we have rewritten the strict inequality that constrains the deployment of VNF 1 as a non-strict one, to guarantee that the related set is closed.}: 
\begin{equation}
d_{c} x_{s,1}+d_{e}(1-x_{s,1})\leq v \cdot b_{s,1}.
\end{equation}

In addition, when $d_{e,c}<v\cdot f_{s,1}$ there may be a split between VNFs $1$ and $2$; then, the computational rate required for VNF $1$ when it runs at the central cloud or at the edge cloud is
\begin{equation*}
\begin{array}{ll}
C_{s,1}^{c+}=\max\left\{\frac{\lambda_{s,1}}{f_{s,1}-\frac{ d_{e,c}}{v}},C_{s,1}^c\right\}\\ 
C_{s,1}^{e+}=\max\left\{\frac{\lambda_{s,1}}{f_{s,1}-\frac{d_{e,c}}{v}},C_{s,1}^e\right\}.
\end{array}
\end{equation*}
Then, we use $\Delta C_{s,1}^{c+}=C_{s,1}^{c+}-C^c_{s,1}$ and $\Delta C_{s,1}^{e+}=C_{s,1}^{e+}-C^e_{s,1}$ to indicate the additional resources required by VNF $1$ when VNF $2$ is at a different cloud; accordingly, for VNF $1$, the  computational rate required at the central cloud or at the edge cloud is as follows:
\begin{equation}
\begin{array}{ll}
\bar{R}^{\mbox{\scriptsize c}}_{s,1}=C_{s,1}^c x_{s,1} + \Delta C_{s,1}^{c+}\Delta x^{c+}_{s,1} ;\\
\bar{R}^{\mbox{\scriptsize e}}_{s,1}=C_{s,1}^e \left(1-x_{s,1}\right) + \Delta C_{s,1}^{e+} \Delta x^{e+}_{s,1},
\end{array}
\end{equation}
where $\Delta x^{c+}_{s,1}=\max\left\{x_{s,1}-x_{s,2}, 0\right\}$ and $\Delta x^{e+}_{s,1}=\max\left\{-x_{s,1}+x_{s,2}, 0\right\}$ indicate if there is a split between VNF $1$, respectively located at the central or at the edge cloud, and VNF 2.

\subsubsection{ILP reformulation}
Finally, by leveraging on the analysis developed in the previous subsection, we reformulate problem $(3)$ as follows:
\begin{mini!}|s|[2]
{\left\{x_{s,n}\right\}}{\sum\limits_{s \in \mathcal{S}}\sum\limits_{n \in \mathcal{N}_s} \bar{R}_{s,n}^c+\bar{R}_{s,n}^e}{}{}\label{P2:dim}
\addConstraint{\sum\limits_{s \in \mathcal{S}}\sum\limits_{n \in \mathcal{N}_s} \bar{R}_{s,n}^c}{\leq C^c}{}\label{P2:dim_con1}
\addConstraint{\sum\limits_{s \in \mathcal{S}}\sum\limits_{n \in \mathcal{N}_s} \bar{R}_{s,n}^e}{\leq C^e}{}\label{P2:dim_con2}
\addConstraint{\hspace{-0.25cm} d_{e,c}\left|x_{s,n} - x_{s,n+1}\right|}{\leq v\cdot f_{s,n},\;s \in \mathcal{S}, n\in\mathcal{N}_s\setminus \left\{N_s\right\}}{}\label{P2:dim_con3}
\addConstraint{d_{e,c}\left|x_{s,n} - x_{s,n-1}\right|}{\leq v \cdot b_{s,n},\; s \in \mathcal{S}, n\in\mathcal{N}_s\setminus \left\{1\right\}}{}\label{P2:dim_con4}
\addConstraint{d_{c} x_{s,1}+d_{e}(1-x_{s,1})}{\leq v \cdot b_{s,1}, \;s \in \mathcal{S}}{}\label{P2:dim_con5}
\addConstraint{x_{s,n}}{\in \{0,1\},\;s \in \mathcal{S}, n \in \mathcal{N}_s}{}\label{P2:dim_con6}.
 \end{mini!}
Note that deriving the expressions of the required computational rates, the latency constraints \eqref{P:dim_con1} and \eqref{P:dim_con2} have been reformulated as functional split constraints \eqref{P2:dim_con3}-\eqref{P2:dim_con5}. As mentioned in the previous subsection, they limit the VNF deployment such that the allocated computational rates $\bar{R}_{s,n}^c$ and $\bar{R}_{s,n}^e$ satisfy the VNF latency requirements.

Although problem $(12)$ is equivalent to problem $(3)$, it can be more efficiently solved with a standard solver as Gurobi \cite{gurobi}. First, the above problem has fewer variables than problem $(3)$. Moreover, the functional split constraints explicitly limit the decision space of $\left\{x_{s,n}\right\}$ to the values that satisfy the latency requirements. Finally, the constraints involving the
absolute value operators can be replaced by linear constraints through simple manipulations \cite{boyd2004convex}.

\section{Extension to the multiple-edge cloud case}
\label{multi-edge}
We now analyze the optimal VNF deployment in a hybrid cloud infrastructure composed of multiple edge clouds and one central cloud.
In this case, depending on the resource availability and network constraints, each VNF can be deployed either at the central cloud or at one of the multiple edge clouds. Recall that we use $\mathcal{K}=\left\{0, 1, 2,\ldots, K\right\}$ to denote the set of nodes in the hybrid cloud infrastructure; then, we define a new VNF association variable as $x_{s,n}^k$, such that $x_{s,n}^k=1$ if VNF $n \in \mathcal{N}_s$ of NF chain $s\in \mathcal{S}$ is associated with cloud node $k \in \mathcal{K}$ and $x_{s,n}^k=0$ otherwise. As in Section \ref{twoclouds}, we use the special structures of our model, especially the per VNF latency constraints, to characterize first the computational resources required by VNF $n\in \mathcal{N}_s \backslash\{1,N_s\}$; then, we analyze the special cases related to VNFs $N_s$ and $1$.

Let us consider the interactions in NF chain $s\in\mathcal{S}$ between VNF $n\in\mathcal{N}_s \backslash\{1,N_s\}$ and its neighbouring VNFs $n+1$ and $n-1$; the minimum computational rate required at cloud $k$ to meet the requirements of VNF $n$ is:
  $$C_{s,n} = \frac{\lambda_{s,n}}{\min\left\{f_{s,n}, b_{s,n}\right\}},$$
  when VNFs $n$, $n+1$, and $n-1$ are co-located at cloud $k$.
 
 In contrast, when there is a split between VNF $n$ and one of its neighbouring VNFs $n+1$ and $n-1$, i.e., $n$ is processed by cloud $k$ and $n+1$ or $n-1$ by cloud $j\neq k$, the computational rate required by VNF $n$ increases to compensate the communication delay between cloud nodes $k$ and $j$ as follows:
 $$C_{s,n}^{(k,j)^+}=\max\left\{\frac{\lambda_{s,n}}{f_{s,n}-\frac{d_{k,j}}{v}},C_{s,n}\right\},~C_{s,n}^{(k,j)^-}=\max\left\{\frac{\lambda_{s,n}}{b_{s,n}-\frac{ d_{k,j}}{v}},C_{s,n}\right\},$$
where $d_{k,j}$ is the distance between nodes $k$ and $j$. Note that the above expressions highlight that there may be a split between VNFs $n$ and $n+1$ or $n-1$ only if the associated communication delay is below the VNF latency constraint, i.e.,  $d_{k,j}< v \cdot f_{s,n}$ and $d_{k,j}< v \cdot b_{s,n}$, respectively. 

Let us now introduce the following notations:
$$\Delta C_{s,n}^{(k,j)^+} =C_{s,n}^{(k,j)^+} - C_{s,n};~
\Delta x^{(k,j)^+}_{s,n}=x_{s,n}^k+x_{s,n+1}^j-1, s \in \mathcal{S}, n\in\mathcal{N}_s\setminus \left\{N_s\right\}, k\neq j \in \mathcal{K}, $$
$$\Delta C_{s,n}^{(k,j)^-} =C_{s,n}^{(k,j)^-} - C_{s,n};~
\Delta x^{(k,j)^-}_{s,n}=x_{s,n}^k+x_{s,n-1}^j-1, s \in \mathcal{S}, n\in\mathcal{N}_s\setminus \left\{1\right\}, k\neq j \in \mathcal{K}.$$ 

In the above, $\Delta C_{s,n}^{(k,j)^+}$ and $\Delta C_{s,n}^{(k,j)^-}$ denote the additional rates, with respect to $C_{s,n}$, required by VNF $n$ deployed at cloud $k$ when VNF $n+1$ and VNF $n-1$ are at cloud $j \neq k$, respectively. 
 {$\Delta x^{(k,j)^+}_{s,n}$ indicates whether there is a functional split between VNF $n$ located at cloud $k$ and its neighbouring VNF $n+1$ deployed at cloud $j \neq k$; specifically, $\Delta x^{(k,j)^+}_{s,n}=0$ if VNFs $n$ and $n+1$ are both located at cloud $k$ and $\Delta x^{(k,j)^+}_{s,n}=1$ otherwise. Similarly, $\Delta x^{(k,j)^-}_{s,n}$ specifies if there is a functional split between VNF $n$ located at cloud $k$ and VNF $n-1$ deployed at cloud $j \neq k$, i.e., $\Delta x^{(k,j)^-}_{s,n}=0$ if VNFs $n$ and $n-1$ are both located at cloud $k$ and $\Delta x^{(k,j)^-}_{s,n}=1$ otherwise.}

Using the above notations, the previously discussed functional split constraints can be now explicitly described as follows: 
\begin{equation*}
    d_{k,j}\max\left\{\Delta x^{(k,j)^+}_{s,n},0\right\}\leq v\cdot f_{s,n} \mbox{ and }
  d_{k,j}\max\left\{\Delta x^{(k,j)^-}_{s,n},0\right\}\leq v \cdot b_{s,n}. 
\end{equation*}

\begin{figure*}[!t]
\normalsize
\setcounter{equation}{12}
\small \begin{equation}
 \label{eq:comp_rate}
{ \bar{R}^k_{s,n}=
C_{s,n}x_{s,n}^k+\max\left\{\max\left\{\max\limits_{j\neq k}\left\{\Delta C^{(k,j)^+}_n\Delta x^{(k,j)^+}_n\right\},0\right\},\max\left\{\max\limits_{j\neq k}\left\{\Delta C^{(k,j)^-}_n\Delta x^{(k,j)^-}_n\right\},0\right\} \right\}.}
\end{equation}
\normalsize
\setcounter{equation}{13}
\hrulefill
\vspace*{4pt}
\end{figure*}

Therefore, for each VNF $n\in\mathcal{N}_s \backslash\{1,N_s\}$, the computational rate required from cloud node $k \in \mathcal{K}$ is described in \eqref{eq:comp_rate} at the top of the page.


In the above expression, the first term corresponds to the minimum computational rate required by VNF $n$ if it is co-located with VNFs $n+1$ and $n-1$; moreover, the second term denotes the additional computational rate required in case of the functional split. Note that, when VNFs $n+1$ and $n-1$ are processed at the clouds $j_1$ and $j_2$, which are different from cloud $k$ where VNF $n$ is running, the additional computational rate depends on the most stringent constraint between $f_{s,n}-d_{k,j_1}/v$ and $b_{s,n}-d_{k,j_2}/v$ and this corresponds to the outermost $\max$ operator in \eqref{eq:comp_rate}. It is also worth mentioning that $0$ in \eqref{eq:comp_rate} corresponds to the trivial case where $j$ is equal to $k.$

We now deal with the special case of VNF $N_s$, where we have only one possible split to consider, i.e., with VNF $N_s-1$. Then, the minimum computational rate that VNF $N_s$ requires (when $N_s$ and $N_s-1$ are co-located) is as follows:
   $$C_{s,N_s} = \frac{\lambda_{s,N_s}}{\min\left\{f_{s,N_s}, b_{s,N_s}\right\}}.$$ In contrast, if $d_{k,j}<v \cdot b_{s,N_s}$, there may be a split between the two functions, i.e., VNF $N_s$ is processed at node $k$ and VNF $N_s-1$ at node $j\neq k$. Accordingly, the computational rate required at cloud $k$ to process VNF $N_s$ may increase, and it is computed as follows:
   $$C_{s,N_s}^{(k,j)^-}=\max\left\{\frac{\lambda_{s,N_s}}{b_{s,N_s}-\frac{d_{k,j}}{v}},C_{s,N_s}\right\}.$$
Therefore, for VNF $N_s$, the computational rate required from cloud $k \in \mathcal{K}$ is
 \begin{equation}
\bar{R}^k_{s,N_s}=C_{s,N_s} x_{s,N_s}^k+
\max\left\{\max\limits_{j\neq k}\left\{\Delta C^{(k,j)^-}_{s,N_s}\Delta x^{(k,j)^-}_{s,N_s}\right\},0\right\}.
 \end{equation}

Likewise, for the special case related to VNF $1$, the minimum computational rate required when it is co-located with VNF $2$ at cloud $k$ is as follows:
$$C_{s,1}^k=\max\left\{\frac{\lambda_{s,1}}{f_{s,1}},\frac{\lambda_{s,1}}{b_{s,1}-\frac{ d_{k}}{v}}\right\},$$
where $d_{k}$ is the distance between cloud $k$ and the RRH, where the RF function of chain $s$ is deployed. 
In the previous expression, we can observe that VNF $1$ can be deployed at cloud $k$, only if the delay to communicate with the associated RRH is below the threshold $b_{s,1}$. This constraint can be expressed as follows:
\begin{equation*}
    d_{k}x_{s,1}^k\leq v \cdot b_{s,1}.
\end{equation*}
In addition, the computational rate required by VNF $1$ when it is deployed at cloud $k$ and VNF $2$ run at cloud $j \neq k$ can be computed as
$$C_{s,1}^{(k,j)^+}=\max\left\{\frac{\lambda_{s,1}}{f_{s,1}-\frac{ d_{k,j}}{v}},C_{s,1}^k\right\}.$$
Therefore, the overall computational rate required from cloud node $k \in \mathcal{K}$ for running VNF $1$ is computed as follows:
 \begin{equation}
\bar{R}^k_{s,1}=C_{s,1}^k x_{s,1}^k+
\max\left\{\max\limits_{j\neq k}\left\{\Delta C^{(k,j)^+}_{s,1}\Delta x^{(k,j)^+}_{s,1}\right\},0\right\}.
 \end{equation}

Based on the above analysis, the problem of minimizing the overall computational rate needed to run $S$ VNF chains in a hybrid infrastructure with multiple edge clouds is described as follows:
\begin{mini!}|s|[2]
{\left\{x_{s,n}^k\right\}}{\sum\limits_{k \in \mathcal{K}}\sum\limits_{s \in \mathcal{S}}\sum\limits_{n \in \mathcal{N}_s} \bar{R}_{s,n}^k}{}{}\label{P3:dim}
\addConstraint{\sum\limits_{s \in \mathcal{S}}\sum\limits_{n \in \mathcal{N}_s} \bar{R}_{s,n}^k}{\leq C^k, \;k \in\mathcal{K}}{}\label{P3:dim_con1}
\addConstraint{d_{k,j}\max\left\{\Delta x^{(k,j)^+}_{s,n},0\right\}}{\leq v\cdot f_{s,n}, \; s \in \mathcal{S}, \;n\in\mathcal{N}_s\setminus \left\{N_s\right\}, \;k\neq j \in\mathcal{K}}{ }\label{P3:dim_con2}
\addConstraint{d_{k,j}\max\left\{\Delta x^{(k,j)^-}_n,0\right\}}{\leq v \cdot b_{s,n}, \; s \in \mathcal{S}, \;n\in\mathcal{N}_s\setminus \left\{1\right\}, \;k\neq j \in\mathcal{K}}{}\label{P3:dim_con3}
\addConstraint{d_{k}x_{s,1}^k}{\leq v \cdot b_{s,1}, \; s \in \mathcal{S}, \;k \in\mathcal{K}}{}\label{P3:dim_con4}
\addConstraint{\sum\limits_{k \in \mathcal{K}} x_{s,n}^k}{=1, \; s \in \mathcal{S}, \;n\in\mathcal{N}_s}{}\label{P3:dim_con5}
\addConstraint{x_{s,n}^k}{\in \{0,1\},  \; s \in \mathcal{S}, \; n \in \mathcal{N}_s, \;k \in\mathcal{K}}{}\label{P3:dim_con6},
 \end{mini!}
 where \eqref{P3:dim_con5} ensures that each VNF is effectively deployed at one and only one of the clouds in the hybrid infrastructure. Note that the constraints including the $max$ operator can be easily replaced by linear constraints \cite{boyd2004convex}.
Therefore, the above problem is still an ILP, although it is more complex than problem (12), mainly due to the higher number of variables. Specifically, assuming that $N_s=N$, $\forall \; s \in \mathcal{S}$, the number of variables is $(K+1)\cdot S\cdot N$ and the number of constraints is in the order of $(NS(K+1))^2$.
An ILP is generally an NP-hard problem and the {worst}-case complexity of globally solving it is $2^{NS(K+1)}\times O((NS(K+1))^2)$, i.e., it increases exponentially with the total number of variables, as finding the optimal solution may require an exhaustive enumeration. Accordingly, to find efficient slice deployment solutions in scenarios with a large set of slices and/or clouds, we propose a low-complexity heuristic based on the special structure of problem (16) in the next section.

\section{A Computationally Efficient Heuristic}
\label{heuristic}
The heuristic presented in this section attempts to find a computational efficient solution in scenarios with a large number of variables. To do so, we exploit the analysis detailed in Section \ref{multi-edge}. More specifically, the proposed heuristic is based on the finding that fully deploying a VNF chain in a single cloud likely lowers the required computational rate, as it does not introduce any communication delay. In addition, we take advantage of the similarity of our ILP with the so-called bin packing problem, and thus we use in the proposed heuristic a \textit{best fit decreasing} strategy \cite{Knapsack}, whose performance are known to be relatively close to the optimal one for the offline bin packing. In the best fit decreasing strategy, the best fit for a new request is the feasible cloud with the smallest available computational capacity (which is likely to be an edge cloud), to limit the amount of resources which may be left unused. Moreover, since dealing with larger computational requests is more difficult, the best fit decreasing scheme sorts the requests in a descending order and attempts to place the slice with the largest computational request first.

Recall that we use $\mathcal{S}$ and $\mathcal{K}$ to denote the set of NF chain requests and the set of cloud nodes in the hybrid cloud infrastructure, respectively  (see Section \ref{sec:format}).
Accordingly, let $\mathcal{S}_{\mbox{\scriptsize R}}:= \mbox{sort}(\mathcal{S}, \mbox{desc.}, \sum\limits_{n \in \mathcal{N}_s} C_{s,n})$ and $\mathcal{K}_{\mbox{\scriptsize R}}:= \mbox{sort}(\mathcal{K}, \mbox{asc.}, C^k)$ denote the sets generated by sorting the elements of $\mathcal{S}$ and $\mathcal{K}$ respectively in the descending and ascending order according to the minimum computational rate required by the VNF chain and the cloud computational capacity.
The proposed `Best Fit with IteRative Split Trial' (B-FIRST) strategy (whose pseudo-code is shown in Algorithm \ref{H1} below) manages one VNF chain request at each iteration; when there is no cloud with sufficient computational resources to deploy the overall chain, the algorithm selects, within the functional split options, the one that enables to accept the VNF chain request with minimum additional computational resources. When a feasible split does not exist or the VNF chain request can be fulfilled, the algorithm attempts to deploy the second most demanding chain, and so on. The algorithm terminates when all VNF chains have been tested.

\begin{algorithm}
\caption{B-FIRST\label{H1}}
\SetAlgoLined
\KwData{Input: $C_{s,n}$, $\bar{R}^k_{s,n}$, and $C^k$, $s\in \mathcal{S}$, $n \in \mathcal{N}_s$, $k \in \mathcal{K}$} 
\KwResult{$\left\{x_{s,n}^k\right\}$, $s\in \mathcal{S}$, $n \in \mathcal{N}_s$, $k \in \mathcal{K}$}

$\mathcal{S}_{\mbox{\scriptsize R}}:= \mbox{sort}(\mathcal{S}, \mbox{desc.}, C_{s,n});
\mathcal{K}_{\mbox{\scriptsize R}}:= \mbox{sort}(\mathcal{K}, \mbox{asc.}, C^k)$\;

\For{each {VNF chain} $s$ $\in$ $\mathcal{S}_{\mbox{\scriptsize R}}$}
{\For{each cloud $k$ $\in$ $\mathcal{K}_{\mbox{\scriptsize R}}$}
{\If{$\sum\limits_{n \in \mathcal{N}_s} \bar{R}_{s,n}^k\leq C^k$}
{$x_{s,n}^k=1, \; n \in \mathcal{N}_s$; $C^k\leftarrow C^k-\sum\limits_{n \in \mathcal{N}_s} \bar{R}_{s,n}^k$;$~\mathcal{K}_{\mbox{\scriptsize R}}\leftarrow \mbox{sort}(\mathcal{K}, \mbox{asc.}, C^k)$\;
}
} 
\If{$\sum_{k \in \mathcal{K}}\sum_{n \in \mathcal{N}_s} x_{s,n}^k\neq N_s$}
{\For{each cloud $k$ $\in$ $\mathcal{K}_{\mbox{\scriptsize R}}$}
{\For{each cloud $j$ $\in$ $\mathcal{K}_{\mbox{\scriptsize R}}$ \mbox{with} $j\neq k$}
{
Find split $p$ between clouds $k$ and $j$ with the smallest resource requirements\;
\If{$\sum\limits_{n \leq p} \bar{R}_{s,n}^k\leq C^k \mbox{and} \sum\limits_{n >p} \bar{R}_{s,n}^j\leq C^j$}
{$x_{s,n}^k=1, \; \mbox{if} \; n \leq p$; $C^k\leftarrow C^k-\sum\limits_{n \leq p} \bar{R}_{s,n}^k$\;
$x_{s,n}^j=1, \; \mbox{if} \; n > p$; $C^j\leftarrow C^j-\sum\limits_{n > p} \bar{R}_{s,n}^k$\;$\mathcal{K}_{\mbox{\scriptsize R}}\leftarrow \mbox{sort}(\mathcal{K}, \mbox{asc.}, C^k)$\;
}
}}}}
\end{algorithm}

Note that, in contrast to problem (16) formulated in Section \ref{multi-edge}, the heuristic can select at most one split within the cloud infrastructure, which limits the system flexibility but also the solution complexity. In fact, the worst-case complexity of B-FIRST is in the order of $(K+1)^2 \sum\limits_{s \in \mathcal{S}} N_s$. In future works, we will investigate how to extend B-FIRST to enable multiple splits with a limited complexity. 

\section{Simulation Results}
\label{sec:simulation}
To assess the proposed VNF deployment framework, we consider {an urban macro scenario where cells are deployed according to a hexagonal layout with an inter-site distance of 500 m \cite{3GPP38913}. The RRHs in the network} provide radio access for multiple slices: eMBB, mMTC, and two types of URLLC: URLLC 1 and URLLC 2. In particular, the mMTC services are characterized by loose latency constraints and low throughput {(small bandwidth and low spectral efficiency)} requirements; the eMBB services have large throughput {(large bandwidth and high spectral efficiency)} and intermediate latency constraints. URLLC 1 models factory automation services and it has very tight latency constraint, {high spectral efficiency requirement} but relaxed bandwidth demand; in contrast, URLLC 2 characterizes services such as augmented reality, and it {requires} large bandwidth{, high spectral efficiency,} and low latency. The MCS indices and the number of RBs per VNF chain {assumed to represent to these requirements} are given in Table \ref{tab:Simulation parameters} and they define the service computational resource demands according to the model in \eqref{comp_req}. 
Each of these {VNF chains} is composed by $N_s=8$ VNFs as described in Fig. \ref{fig:Functional split}.
As already mentioned, the latency constraints of each VNF depend on the timing requirements of the interfaces with its neighbouring VNFs \cite{maeder2014} and the overall service latency constraint. The considered requirements $\left\{b_{s,n}\right\}$ for each type of chain $s$ and VNF $n$ are shown in Table \ref{Lat_Req}. In this work, without loss of generality, we set $f_{s,n}=b_{s,n+1}$.
The values of the other parameters needed for the computational resource model in \eqref{comp_req} can be found in \cite{Khatibi2018}. Moreover, unless otherwise specified, we consider that the set of requested VNF chains $\mathcal{S}$ includes only one mMTC (due to its loose service requirements) and is composed of an equal number of eMBB and URLLC services; for example, if $S=7$, the request set includes one mMTC chain, two eMBB chains, two chains of URLLC 1, and two chains of URLLC 2. The mMTC chain is associated with the RRH at the central macro cell; however, the other VNF chains are randomly associated with the network RRHs. Regarding the cloud infrastructure, we consider that an Intel Xeon Platinum 8180M Processor ($f_{\mbox{\scriptsize CPU}}=$2.50 GHz) is used at the central cloud and Intel Xeon Silver 4114T Processors ($f_{\mbox{\scriptsize CPU}}=$2.20 GHz) are deployed at the edge clouds \cite{Intel}.
To conclude, we solve problems (12) and (16) using Gurobi \cite{gurobi}, which uses a branch-and-cut algorithm for ILP problems{, and we set a time limit of 600 seconds to find the optimal solution of any instance of these problems}.  

\begin{table}[!h]
\caption{RBs and downlink and uplink MCS indices for various types of services.}
\label{tab:Simulation parameters}
\small
\centering
  \begin{tabular}{|l||c|c|c|c|} 
  \hline
&eMBB  &  mMTC & URLLC 1 & URLLC 2 \\
  \hline 
  $RB_{s}$ & 250 & 5 &25 &500\\
  \hline 
  $i_{s,\mbox{\scriptsize DL}}$ & 27 & 13 &27 &27\\
  \hline 
$i_{s,\mbox{\scriptsize UL}}$ & 16 & 8 &16 &16\\
   \hline 
  \end{tabular}
\end{table}
\begin{table*}[!h]
\caption{Latency constraints for VNFs in different services \cite{maeder2014}.}
\label{Lat_Req}
\small
\centering
  \begin{tabular}{|c||c|c|c|c|c|c|c|c|c|c|c|c|c|c|} 
  \hline
   \scriptsize 
   &$n=1$&$ n=2$& $ n=3$& $ n=4$& $ n=5$& $n=6$& $n=7$&$n=8$& {Total}\\
  \hline 
$b_{\mbox{\scriptsize eMBB},n}$ [ms] & 1 &3&3&3 &{22.5}  &{22.5} & {22.5} & {22.5} &{100} \\
  \hline 
$b_{\mbox{\scriptsize mMTC},n}$ [ms] & 10& 10& 10& 10 &200  &500& $10^4$& $2\cdot10^3$ &{12.74$\cdot10^3$}\\
   \hline 
$b_{\mbox{\scriptsize URLLC 1},n}$ [ms]  &0.2&0.2&0.2&0.2 &0.2  &0.2& $0.2$& $0.2$&{1.6}\\
  \hline 
$b_{\mbox{\scriptsize URLLC 2},n}$  [ms] &0.5&0.5&0.5 &0.5 &0.5  &0.5& $0.5$& $0.5$&{4} \\
  \hline 
  \end{tabular}
\end{table*}

\subsection{Two-Cloud Hybrid Infrastructure}
In this section, we focus on a simple cloud infrastructure composed of a central cloud and a single edge cloud {located in the central macro cell}. The goal is twofold: first, we aim to assess the advantage of the hybrid architecture with respect to a classical C-RAN solution that only includes a central cloud; then, we target to highlight the advantage of our optimal VNF deployment scheme over less flexible solutions.

\begin{figure}
\centering
\includegraphics[width = 9cm]{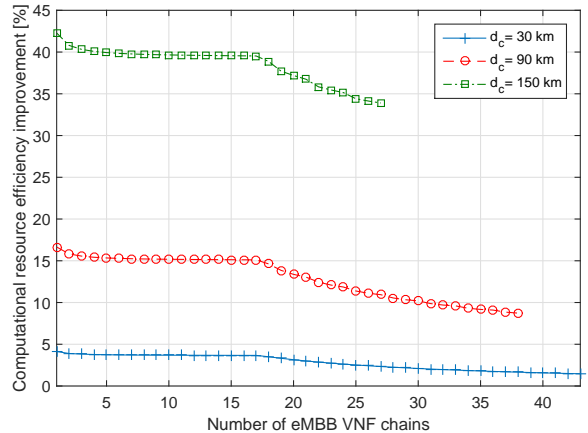}
\caption{Computational resource efficiency improvement of the hybrid cloud infrastructure with respect to a C-RAN solution as a function of the number of accepted eMBB VNF chains.}
\label{fig1}
\end{figure}

{We set the cloud computational capacity $C^c$ to be 13440 GFLOPS/s in the classical C-RAN architecture; in the hybrid architecture, the central cloud has two thirds of the overall capacity, i.e., 8960 GFLOPS/s, while the rest is available at the edge cloud, i.e., $C^e$=4480 GFLOPS/s.}
Fig. \ref{fig1} shows the computational resource efficiency improvement provided by the hybrid cloud infrastructure with respect to a C-RAN architecture, composed only of a central cloud, as a function of the number of deployed VNF chains and for different distances of the central cloud from the central macro cell. This metric measures the reduction of computational resources led by the hybrid infrastructure with respect to the C-RAN architecture, when serving a given set of VNF chains. In this simulation, we focus on the optimal deployment of chains only related to the eMBB service, in order to clearly evaluate the advantages of the hybrid architecture over the C-RAN solution. Note that including services such as URLLC in the analysis likely increases the gain of the hybrid architecture, as their latency constraints may not be satisfied when the distance between the associated central cloud and the RRH is large. 

First, we observe from Fig. \ref{fig1} that, as expected, the larger the distance of the central cloud from the access network, the larger the gain of the hybrid infrastructure. When the distance is equal to 30 km, having an edge cloud leads to a limited improvement; in fact, we measure that the hybrid architecture requires only 5$\%$ less computational rate as compared to the C-RAN with only a central cloud. However, up to 17$\%$ and 43$\%$ of reduction in terms of resource usage are achieved when the central cloud is located at 90 km and 150 km from the access network, respectively. These improvements come from the relation between communication delay and computational resources in a cloud infrastructure: deploying a VNF at a distant cloud increases its experienced communication latency, which has to be compensated by increasing the computational rate in order to reduce the processing latency and satisfy the VNF latency constraints. 
For a given distance, the achieved improvement slowly varies when the number of chains is low; then, beyond a given number of chains (16 in our results), the edge cloud starts to saturate and the central cloud is used also in the hybrid infrastructure, which notably decreases the computational resource efficiency gain. 

It is worth noticing that the improvement in terms of computational resource usage provided by the hybrid cloud infrastructure paves the way for a larger number of chains that can be supported with respect to a C-RAN architecture with the same computational capacity. In fact, although the C-RAN solution deploys up to 38 and 27 chains when the central cloud is located at 90 km and 150 km, far from the access network, in the same condition, the hybrid infrastructure enables to serve up to 42 and 39 chains, which correspond to a gain of 11$\%$ and 44$\%$ in terms of the number of deployed chains.

\begin{figure}
\centering
    \subfloat[]
{\includegraphics[width = 9cm]{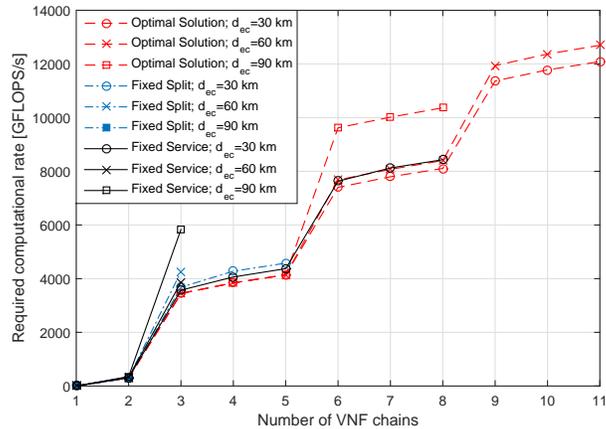}
    \label{fig2}}
    \hfill
    \subfloat[]
    {\includegraphics[width = 9cm]{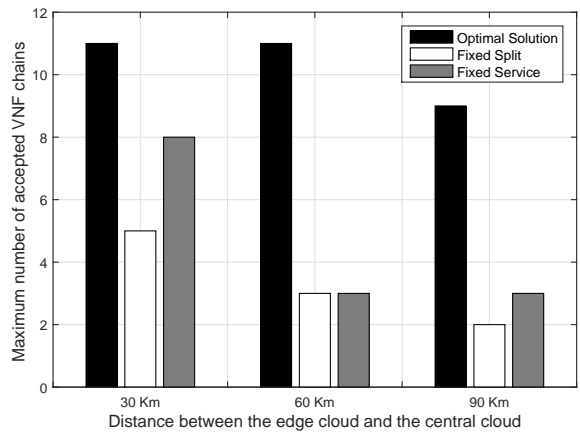}
    \label{fig2b}}
    \caption{(a) Required computational rate of different VNF deployment schemes versus the number of deployed VNF chains. (b) Maximum number of accepted VNF chains as a function of the distance between the edge cloud and the central cloud. $C^c$=8960 GFLOPS/s and $C^e$=4480 GFLOPS/s.}
\end{figure}

 Now, we consider the case of mixed VNF chains, composed of mMTC, eMMB, and two URLLC services described in the previous subsection. We compare the performance of the optimal deployment scheme with two simple solutions denoted as \textit{fixed split} and \textit{fixed service}. In the fixed split solution, the VNFs of each chain, independently of the type of services, are split in the same manner. Specifically, the VNFs up to the lower MAC (see Fig. \ref{fig:Functional split}), which have tighter latency and computational requirements, are deployed at the edge cloud, while the other VNFs are instantiated at the central cloud. In contrast, in the fixed service scheme, the {mMTC}, eMBB, and URLLC 1 chains are always deployed at the central cloud, while only the URLLC 2 chains (which has the stringest latency constraints) are allocated to the edge cloud.
 
 Fig.~\subref*{fig2} shows the overall computational rates required by the hybrid infrastructure, when using different VNF deployment schemes, as a function of the number of VNF chains.
 Dashed, solid, and dotted-dashed lines respectively represent the optimal solution, the fixed service scheme, and the fixed split scheme, and circle marked, cross marked, and square marked lines describe the performance when the central cloud is located at 30, 60, and 90 km from the edge cloud.
First of all, we can see as for a fixed number of accepted VNF chains, the reduction in terms of required computational rate of the optimal solution with respect to the fixed service and the fixed split schemes increases with the distance between the central cloud and the edge cloud: up to 5$\%$ and 10$\%$ for $d_{e,c}=$ 30 km and up to 11$\%$ and 19$\%$ for $d_{e,c}=$ 60 km. For $d_{e,c}=$ 90 km, we achieve up to 41$\%$ gain with respect to the fixed service scheme; however, we cannot measure appreciable gains with respect to the fixed split scheme, since it fails to deploy more than two chains due to the large distance between the two clouds. When the number of VNF chains to deploy is very low or the central cloud is located near the access network, the static schemes have similar performance as that of the optimal solution; however, in the other scenarios, either they require a much larger computational rate or they fail to find a distribution of the resources that satisfies the service requirements. In sharp contrast, the optimal scheme adapts the functional split at each accepted VNF chain, and when a new request arrives it is able to rearrange the available resources such that the system performance is optimized.

In fact, we can observe from Fig. \subref*{fig2b} that the proposed optimal scheme greatly increases the number of chains that can be successfully deployed with respect to the two static solutions, even when the central cloud is located near the edge cloud (and the macro cell network).
Specifically, for $d_{e,c}=$ 30 km, the optimal solution supports 11 VNF chains, while the fixed service and the fixed split schemes support 8 and 5 chains, with a gain of 37.5$\%$ and 120$\%$ in terms of the number of deployed services. These gains further increase when the distance between the central cloud and the edge cloud becomes larger: the optimal solution leads to a 266$\%$ gain with respect to both the two static schemes in terms of the number of served chains when $d_{e,c}=$ 60 km. Moreover, when $d_{e,c}=$ 90 km, it gains up to 166$\%$ and 300$\%$, with respect to the fixed service and the fixed split schemes. {
In addition, it is worth to highlight that the fixed service scheme always results in better performance than the fixed split approach. In fact, even when they both enable to deploy the same maximum number of VNF chains (see the results in Fig. \subref*{fig2b}),  Fig.~\subref*{fig2} shows that the fixed service scheme requires less computational rate than the fixed split solution for achieving this result.}
Overall, we can conclude that, when the number of VNF chains in the system or the distance between the central cloud and the edge cloud $d_{e,c}$ increase, the proposed scheme brings the desired flexibility to balance the cloud load (i.e., moving VNFs from one cloud to another) to make computational resources available for the chains with more stringent requirements. In contrast, the static schemes lack of such flexibility and lead to a limited performance.

\subsection{Multi-Cloud Hybrid Infrastructure}

In this section we analyze the performance of the optimal scheme for deploying VNF chains when each macro cell of the network is equipped with an edge cloud node. {Note that, in our setup, the distance between two edge clouds is in the range between 0.5 and 1 Km, which is much lower than the distance between the central cloud and a macro cell, and hence it has a limited impact on the system performance. Therefore, in the following, we focus on investigating the impact of the distance between the central cloud and the access network on the system performance.}

\begin{figure}
\centering
    \subfloat[]
{\includegraphics[width = 9cm]{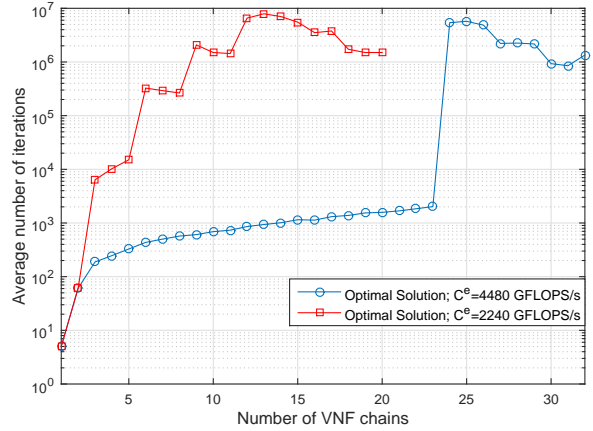}
    \label{Itera}}
    \hfill
    \subfloat[]
    {\includegraphics[width = 9cm]{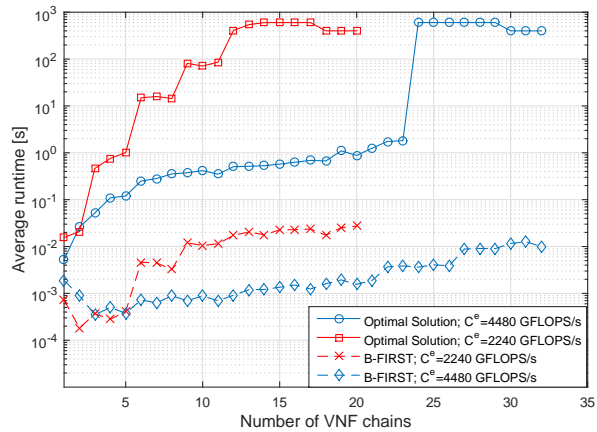}
    \label{Runtime}}
    \caption{(a) Average number of iterations required to find the optimal solution versus the number of VNF chains;  (b) Average runtime of the optimal solution and B-FIRST. $C^c$=8960 GFLOPS/s.}
\end{figure}

First, to highlight the necessity and importance of developing a simple low-complexity heuristic, we assess the complexity of the proposed optimal scheme in the hybrid architecture with multiple edge clouds. To do so, {in Fig. \subref*{Itera},} we report the number of iterations that Gurobi requires to converge to the optimal solution as a function of the number of VNF chains, for two values of edge cloud capacity and averaging over three different values of the distance between the central cloud and the central macro cell, i.e., 30, 60, and 90 km.
As expected, the number of iterations strongly depends on the number of VNF chains, which increases the number of optimization variables and thus the feasible combinations in the deployment problem. 
Very interestingly, comparing the solution complexity of the two cases, we observe that when the edge cloud capacity is small, the required number of iterations increases faster at first but eventually, for both values of edge cloud capacity, it increases up to $10^7$ when the cloud infrastructure is saturating. Such a large number of iterations results in a large delay, which might not be acceptable when adjusting the deployment of VNF chains with tight latency constraints.
{
To further analyze this issue, Fig. \subref*{Runtime} shows the comparison between the average runtime required to find the optimal solution and to complete the B-FIRST algorithm with respect to the number of VNF chains and for two different values of $C^e$. First, we can observe that the average runtime to get the optimal solution rapidly saturates to the time limit of 600 seconds. In contrast, the runtime needed by B-FIRST slowly grows with the number of VNF chains, which highlights that the proposed heuristic is a promising solution when network slices and the associated computational resources have to be rapidly managed according to the system dynamics.}

\begin{figure}
\centering
    \subfloat[]
    {\includegraphics[width = 9cm]{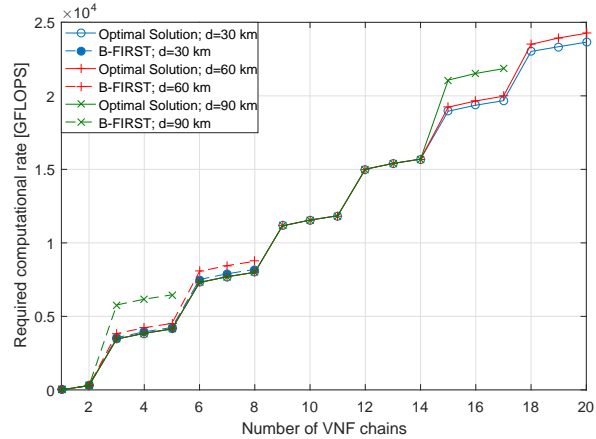}
    \label{fig:resources_MEa}}
    \hfill
    \subfloat[]
    {\includegraphics[width = 9cm]{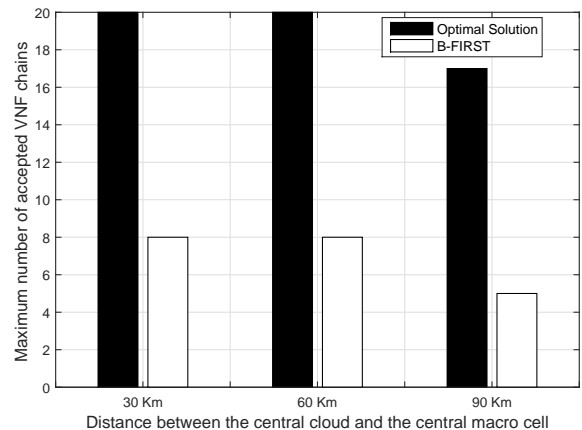}
    \label{fig:number_MEa}}
    \caption{(a) Overall computational rates required by the optimal solution and the proposed heuristic B-FIRST as a function of the number of accepted VNF chains. (b) Maximum number of accepted VNF chains versus the distance between the central cloud and the central macro cell. $C^c$=8960 GFLOPS/s and $C^e$=2240 GFLOPS/s.}
\end{figure}

Fig. \subref*{fig:resources_MEa} describes the computational rates required by the optimal solution and the proposed B-FIRST heuristic with respect to the number of VNF chains. We assume here that the edge clouds and the central cloud have a computational capacity equal to 2240 GFLOPS/s and 8960 GFLOPS/s, respectively. In Fig. \subref*{fig:resources_MEa}, solid and dashed lines correspond to the performance of the optimal solution and the heuristic scheme, respectively. Moreover, circle marked, plus marked, and cross marked lines correspond respectively to the cases where the central cloud is 30, 60, and 90 km far from the central macro cell. 

First, we can notice that, when the central cloud distance is not large (i.e., 60 km and below) and the number of required VNF chains is limited (up to 8), the performance of the heuristic and the optimal scheme is very close; in fact, in this case, both the optimal and the heuristic schemes deploy as many VNF chains as possible at the edge clouds, and when a VNF chain has to be deployed at the central cloud, the additional computational rate required in the system is limited. 
However, when the central cloud is located at 90 km from the macro cell, even for a small number of VNF chains (i.e., 3 to 5), the optimal solution improves of 56$\%$ the computational resource efficiency with respect to the proposed heuristic. In fact, in this case it is necessary to properly select the VNFs to deploy at the central cloud, as those with stringent latency constraints will require a large amount of additional computational resources to compensate for the communication delay introduced by the central cloud.

From Fig. \subref*{fig:number_MEa}, we can clearly see that the optimal solution leads to a notable performance gain with respect to the proposed heuristic in terms of the number of supported VNF chains. Specifically, the optimal solution enables the 5G system to accept up to 20 VNF chains when the central cloud is 30 km and 60 km far from the central macro cell, and up to 17 VNF chains when the central cloud is 90 km far from the macro cell. In contrast, with B-FIRST, the 5G system can only accept up to 8 VNF chains when the central cloud is 30 km and 60 km far from the central macro cell, and up to 5 VNF chains when the central cloud is 90 km far from the macro cell. This gain is due to the ability of the optimal solution to introduce multiple splits in each VNF chain when the system is saturated, which enables an optimal usage of the available computational resources distributed through the cloud architecture. In contrast, imposing a single split in the chain, as in the heuristic, strongly reduces the system performance when the edge cloud capacity is relatively small. 

\begin{figure}
\centering
    \subfloat[]
{\includegraphics[width = 9cm]{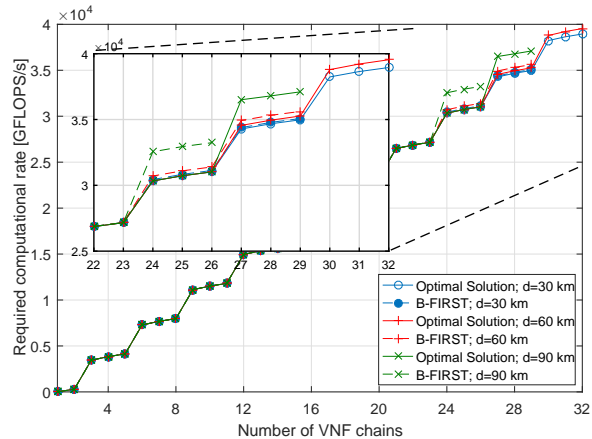}
    \label{fig:resources_MEb}}
    \hfill
    \subfloat[]
    {\includegraphics[width = 9cm]{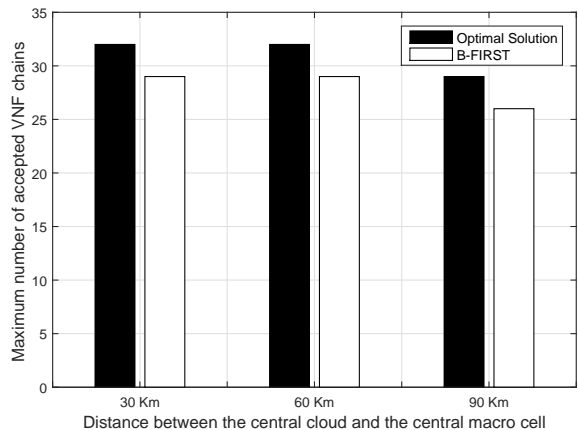}
    \label{fig:number_MEb}}
    \caption{(a) Overall computational rates required by the optimal solution and the proposed B-FIRST heuristic as a function of the number of accepted VNF chains. (b) Maximum number of accepted VNF chains versus the distance between the central cloud and the central macro cell. $C^c$=8960 GFLOPS/s and  $C^e$=4480 GFLOPS/s.}
\end{figure}

We now consider the case where the edge cloud capacity is larger and is set to be 4480 GFLOPS/s.
In contrast to the previous results, the optimal and the heuristic approaches have very similar performance, even when the central cloud is located at the largest considered distance. 
As we show in the zoomed plot inside Fig. \subref*{fig:resources_MEb}, the two approaches have the same trend until the system comes very close to the saturation. In fact, a resource utilization gain is appreciable, i.e., 10$\%$, only when the central cloud is located at a distance of 90 km from the central macro cell and the number of VNF requests is larger than 23.
Also, Fig. \subref*{fig:number_MEb} shows that for a larger number of requests (from 26 to 32), the optimal scheme brings a gain of around 10$\%$ in terms of the number of accepted chains. 
In the scenarios shown in Fig. 7, there is less need for flexibility and complexity as the edge clouds are well dimensioned to satisfy large computational requests, such as those related to the URLLC services, and the proposed heuristic can be used as efficient solution for the VNF deployment in hybrid cloud infrastructures. 
{In contrast, the results in Fig. 6 highlight the need for more complex algorithms offering gains closer to the optimal performance when the edge cloud capacity is limited.}

\begin{figure}
\centering
\includegraphics[width = 9cm]{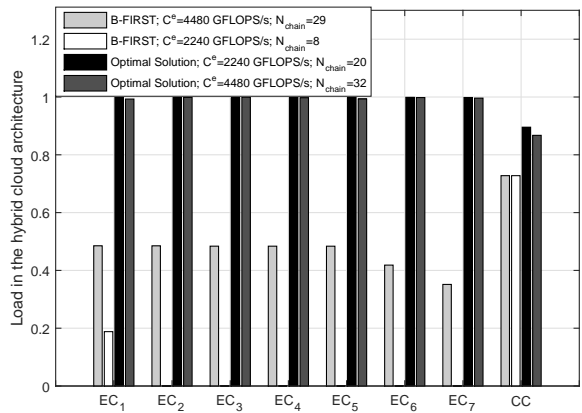}
\caption{Load distribution in the hybrid cloud architecture for the optimal and heuristic solutions; $C^c$=8960 GFLOPS/s; d=30 Km.}
\label{Load}
\end{figure}

To confirm the above analysis, we show in Fig. \ref{Load}, the computational load distribution with the optimal solution and the proposed heuristic, for both cases of edge cloud capacity. Here, we use $EC_j$ to denote the $j$-th edge cloud and $CC$ the central cloud.
We can observe that, for the scenario with edge clouds characterized by large computational capacity ($C^e$=4480 GFLOPS/s), the load distributions of the heuristic and the optimal solutions are very similar, and both approaches achieve a high resource utilization efficiency. In contrast, when the edge clouds have limited computational resources ($C^e$=2240 GFLOPS/s), we can see that, with the heuristic scheme, the central cloud rapidly saturates, while most of the edge cloud resources are not used, which prevents it from accepting further VNF chains. The resources available at the idle edge clouds should be (re)distributed in order to improve the system performance.

\section{Conclusion}
\label{sec:conclusion}
In this paper we have investigated the optimal VNF  deployment problem in a hybrid cloud infrastructure. The problem
is analyzed and formulated as an ILP that can be optimally solved by using standard solvers. Our results highlight the benefit of using a hybrid cloud with respect to a classical C-RAN architecture composed only of a central cloud, in particular for services with tight latency requirements. In addition, we show that the proposed optimal framework is able to fully exploit the resources available at cloud nodes, with small or large computational capacity, by distributing the load in the network and making computational resources available where needed. This significantly improves the resource utilization efficiency and the number of VNF chains that the system can support with respect to static solutions. We have also proposed a simple low-complexity heuristic that performs very close to the optimal solution when the capacity of the hybrid cloud architecture is well dimensioned with respect to the slice computational requests. This highlights the need to investigate optimal cloud design in case of network slicing with heterogeneous requirements. In future works, {we plan to investigate how to deploy our heuristic on a testbed to consider realistic implementation constraints such as orchestration, instantiation, and configuration delays. In addition,} we shall consider dynamic settings, where artificial intelligence based solutions learn the trends of the computational requests and proactively distribute the requests through the clouds, in order to prevent service outage.

\bibliographystyle{IEEEtran}
\bibliography{reference.bib}

\end{document}